\documentstyle[aps,prl,epsfig]{revtex}
\def\cleardoubleemptypage{\clearpage\if@twoside \ifodd\c@page\else
    \thispagestyle{empty}\hbox{}\newpage
    \if@twocolumn\thispagestyle{empty}\hbox{}\newpage\fi\fi\fi}

\newcommand{\be}{\begin{equation}}
\newcommand{\ee}{\end{equation}}
\newcommand{\bea}{\begin{eqnarray}}
\newcommand{\eea}{\end{eqnarray}}

\newcommand{\p}{\partial}
\newcommand{\s}{\sigma}
\newcommand{\rsi}{\rm \sigma}

\newcommand{\la}{\langle}
\newcommand{\ra}{\rangle}
\newcommand{\rd}{\mbox{d}}
\newcommand{\ri}{\mbox{i}}
\newcommand{\re}{\mbox{e}}
\newcommand{\rc}{{\rm c}}
\newcommand{\rs}{{\rm s}}

\newcommand{\rj}{{\rm j}}

\begin{document}

\title{Nonuniversal properties of the single-particle density of states of 
1D system with electron-phonon interactions}
\author{Emiliano Papa\\
{\em{Department of Physics, The University of Texas,
Austin TX 78712}}}
\maketitle
\vspace{0mm}

\address{\rm (Received: )}

\begin{abstract}
\par
We have calculated the single-particle density of states (DOS) for
a model of spinfull Tomonaga-Luttinger liquid  with frequency dependent parameter $K_c$ of the charge 
sector (and $K_s=1$ of spin sector). 
Such frequency dependence may originate from interactions with 
optical phonons. 
DOS exhibits a power-low suppressed asymptotic behaviour near $\omega=0$
 with exponent larger
than one, in agreement with previous results. For larger frequencies, but still
not far away from the origin,
DOS exhibits a peak, the position of which scales as
$[K(0)+K^{-1}(0)]\omega_0$, where $\omega_0$ is the characteristic phonon
frequency. The position of the peak decreases as the phonon's
frequency becomes smaller. It is interesting that the peak of DOS does not
coincide with the characteristic phonon frequency. 
\end{abstract}
\vspace{5mm}

This paper investigates the properties of the single-particle density of states (DOS) at the Fermi surface 
of 1D conductors with electron-electron and electron-phonon interactions. 
Interest in the problem of the electron-phonon 1D systems has been motivated mainly by 
many interesting results on organic conductors and superconductors, conducting polymers
and also photoemission experiments on the properties of the spectral function near the fermi 
surface \cite{blue}, which are known to be influenced strongly by the electron-phonon coupling.  
The electron-phonon interaction in quasi-one dimensional materials has
been studied by different authors, but mostly
for the case of the noninteracting electrons \cite{Heeger,Grunner,Fukuyama,Varga}.
Here the effects of interaction of the charge density 
with small momentum phonons were treated and the direct diagonalization of the Hamiltonian 
could be done.
The combined effects of Coulomb and electron-phonon interactions for phonons with finite 
frequencies was studied in the case of spinless fermions by \cite{Voit and Schulz 0} 
and the spinfull case by the same authors \cite{Voit and Schulz 1}. 
In these papers the effects of dispersionless phonons with momenta $\approx 2k_F$ 
were investigated. In both cases the renormalization group approach was used.
The effort on those papers was concentrated, however, on the effective interaction constants, effective mass due to 
electron-phonon coupling and renormalized velocities for the charge and spin flucuations. 
A possible 
phase diagram (in presence of suitable coupling mechanisms between chains) was represented. 

In this paper we consider the effects of electron-phonon, in addition to
electron-electron interactions, in the framework of Tomonaga-Luttinger
liquid model.
We focus on large momentum transfer scatterings, of $q\approx 2k_F$, $q\approx 4k_F$, 
of electrons by phonons.
The  electron-electron part of the Hamiltonian [Eq.~({\ref{first}) below] 
can be written as a sum
of two independent (commuting) contributions of charge and spin collective
modes. 	To count for the phonons, in the continuous description, 
we separate in fast and slow components the deformation 
field of the lattice sites. We keep only the components near $2k_F$ and $4k_F$  of this field.
Physically the $2k_F$ component interacts with electrons in which one electron hopes from one 
Fermi point to the other. Similarily the $4k_F$ component interacts with electrons and now 
two electron can hope from one Fermi point to the other.     
We integrate out the phonon degrees of freedom and examine the remaining effective 
electron-electron interactions.
The electron-phonon interaction, as observed also in the papers cited above,
affects the charge as well as the spin
sector by renormalizing the coupling constant $g_s$ and the Tomonaga-Luttinger
liquid parameter of the charge sector $K_c$. The study here is confined to the case when 
$g_s$ remains negative, i.e. on the case of gapless spin excitations.
Our approach is similar to the one of Voit and Schulz \cite{Voit and Schulz 0,Voit and Schulz 1}, 
but we differ on the way we treat the retardation 
effects appearing from the presence of the electron-phonon interactions. 
In \cite{Voit and Schulz 0,Voit and Schulz 1} they use a perturbative 
expansion  to study the retardation effects whereas we investigate their effects by studying the 
effective Luttinger liquid to which the electron-phonon interactions give rise to the 
frequency dependend parameter $K_c$ of the charge sector.  
We find the imaginary time single particle Matsubara Green's function and numerically 
produce the single-particle density of states of this unusual Luttinger liquid.

The model we consider is the one-dimensional (1D) Hubbard model coupled to 
phonons.  The lattice effects can be included in this model by making
the hopping integral $t$ dependent on the intercite distance:
\be
t_{ij}\approx t+\frac{1}{2a}\kappa (u_{i}-u_{j})
\quad .
\ee
This has been proposed for the first time by Su, Schriefer and Heeger \cite{Su} to describe the physics of 
conducting polymers.
The Hamiltonian with the above hopping integral has the following form:
\begin{eqnarray}
\label{first}
H= -t\sum_{j,\sigma}\left(c_{j+1,\sigma}^{\dagger}c_{j,\sigma}
+{\rm H.c.}\right)
 +  U\sum_{j }n_{j \uparrow} n_{j \downarrow}
-\frac{1}{2a}\kappa \sum_{j,\sigma}
\left(u_{j}-u_{\rm j+1}\right)
\left(c_{j+1,\sigma}^{\dagger}c_{j,\sigma}
+{\rm H.c.}\right)
+H_{\rm ph}
\ ,
\end{eqnarray}
where $u_{\rm j}$ is dimensionless and $\kappa$ has dimensions of
energy. The $c_{j,\rsi}$ operators are the usual creation and annihilation
operators for the electrons with spin $\sigma$ in the Wannier orbitals
at site $j$ and $n_{\rj,\s}$ is the number
of electrons on site $j$. U is the repulsion of two electrons on the same site.
We consider here the case of the incommensurate band filling i.e. 
$4 k_{\rm F}\not= 2\pi/a$, and therefore we do not consider umklapp scattering. 


Substituting the creation and annihilation operators in terms of the new
continuous fields $R_{\rsi}(x), L_{\rsi}(x)$, 
and introducing the scalar U(1) and vector SU(2) currents 
$J=:L^{\dagger}_{\alpha}L_{\alpha}:$, ${\bf J}=:L^{\dagger}_{\alpha} 
(\vec{\sigma}_{\alpha,\beta}/2) L_{\beta}:$,
the first term of (2) can be written as a sum  $H_1 =  H_1^{U(1)}+H_1^{SU(2)}$. 
Adding to the above part of the Hamiltonian the $U$-interaction term one has
$
H_{\rm Hubb} = H_{\rc} + H_{\rs}
$,
where
\bea
H_{\rc} 
 =  \frac{\pi \bar{v}_{\rc}}{2} \int {\rm d} x
\left(J J + \bar{J} \bar{J} \right) + g_{\rc} \int {\rm d}
x J \bar{J}
\label{ninteen}
%
%
\quad , \quad
H_{\rs} = 
 \frac{2 \pi \bar{v}_{\rs}}{3} \int {\rm d} x
\left({\bf J}{\bf J} + {\bf \bar{J}}{\bf \bar{J}} \right)
+g_{\rs} \int {\rm d} x {\bf J} {\bf \bar{J}}  
\quad ,
\label{canonical0}
\eea
with Fermi velocities $\bar{v}_{\rc}=v_{F} \left(1-Ua/(8 \pi v_{F})\right)$ and
$\bar{v}_{\rs}=v_{F}(1-9 Ua/(8 \pi v_{F}))$ and
coupling constants $g_{\rc}=Ua/2$, $g_{\rs}=-2Ua$, in the charge and
spin sectors respectively.
The coupling constant in the spin sector is negative which  results in 
the spin excitations being gapless (the $g_\rs {\bf J} {\bf
\bar{J}}$ term is marginally irrelevant in the case $g_\rs$ is negative).
Using Abelian bosonization one can restore to $H_\rc$ given by 
(\ref{ninteen}) the
canonical Gaussian form:
\be
H_\rc=\frac{v_\rc}{2} \int {\rm d} x \left[K^{0}_c \Pi^2_\rc+\frac{1}{K^{0}_c}
\left(\partial_x \Phi_\rc \right)^2\right]
\quad ,
\label{canonical}
\ee 
whith $v_\rc$ and $K_\rc^{0}$ depending on the coupling constant $g_\rc$
in the following way: $v_\rc=\bar{v}_\rc(1-g^2_\rc/\pi^2\bar{v}_\rc^2)^{1/2}$ and 
$K^{0}_c= [(1-g_\rc/\pi \bar{v}_\rc)/(1+g_\rc/\pi \bar{v}_\rc)]^{1/2}$.
We will see that the electron phonon interactions lead particularly to strong
renormalization of the parameters of the charge sector. 

The effect of electron-phonon interactions on the dynamics of the electronic system can be easily
calculated since the phononic action is quadratic in the displacement field whereas the 
electron-phonon interaction is a linear function of the lattice displacement (multiplied by the 
electronic staggered charge density). Integrating the displacement fields, in the integrals for the
partition function, one gets in addition to (\ref{canonical0}), an effective interaction between the staggered 
components of the electronic 
charge densities (see for instance \cite{GNT})
\bea
S_{\rm int} = -2\pi v_c \lambda_l\int \rd \tau \rd \tau' \rd x
\left\{\sum_{l=1,2}\int {\rm d}^2 x {\rm d} \tau'
\rho^{*}(2lk_{\rm F},x,\tau')
\left[\delta(\tau'-\tau)  + \frac{1}{2\omega_l} \p_{\tau'}^2 e^{-\omega_l|\tau'-\tau|}\right]
\rho(2lk_{\rm F},x,\tau) \right\} .
\label{S_int}
\eea
where
\bea
\rho(2 k_{\rm F},x) & = & \sum_{\rsi}R^{\dagger}_{\rsi}(x)L_{\rsi}(x)
= a_c \exp\left\{\ri \sqrt{2 \pi}\Phi_c(x) + 2 \ri k_F^{} x \frac{}{}\right\}
\cos\{\sqrt{2\pi}\Phi_{s}^{}(x)\} \quad ,
\label{charge_2kF}  
\eea
\bea 
\label{charge_4kF}
\rho(4 k_{\rm F},x) & = & a_u \exp\left\{\ri \sqrt{8 \pi}\Phi_c(x) + 4 \ri k_F^{} x
\frac{}{}\right\} \quad ,
\label{charge4kF}
\eea
are the $2k_{\rm F}$ and $4k_{\rm F}$-components of the staggered charge density
end $\omega_l=\omega(2lk_{\rm F})$ for $l=1,2$ are the phonon frequencies and 
$\lambda_l=\left[\kappa \sin(l k_{F} a)\right]^{2}/
(\pi v_{\rc} \rho_{l} \omega_{l}^2 a^2)$, where $\rho_l$ is the effective mass of the $l$-th mode.
It is of course known [and can be seen also in the above formula (\ref{S_int})] 
that the electron-electron interaction mediated through phonons
is constituted of two kinds of interactions. First there is an instantaneus attraction effect between
the electronic staggered charge densities.  By observing the formulae for the charge densities 
Eqs.~(\ref{charge_2kF}-\ref{charge_4kF}) one can see that only the $2k_{\rm F}$
component gives a contribution to the instantaneous component of the interaction. The $4k_{\rm F}$
component of the staggered charge density have the particular exponential form 
of complex argument and does not give any contribution. Quantitatively the effect of the 
instanteneous attraction between the $2k_{\rm F}$ components of the staggered charge density
can be evaluated using the identity \cite{identity}. 
This shows that this kind of interaction renormalizes the LL parameters of 
both the charge and the spin channels. In the latter the additional constant will be with positive 
sign and will tend to increase $g_s^{0}$ to a value which we will assume to be still negative.
A positive $g_s$ will lead to a gap in the spin sector, something we are not undertaking to
examine here.

The contribution from the nonlocal term, or the retarded attraction between the staggered 
charge densities, will be discussed a bit longer in the following since they are the factor 
that leads to the frequency dependence of the Luttinger liquid parameter $K_c$ in the charge 
sector. Let us see first the effect of the interaction between the $4k_F$ components of the staggered charge density: 
\bea
\nonumber
& & \, \, \int {\rm d} x \ {\rm d} \tau'
:\rho(4 k_{F},x,\tau) :
\left(\frac{\pi v_\rc \lambda_2}{2 \omega_2}\partial^2_{\tau}
\re^{-\omega_2 |\tau - \tau'|}  \right) 
: \rho(-4 k_{F},x,\tau'):
\\ [2mm]
&=&
 - \int {\rm d} x \ {\rm d} \tau' :\partial_{\tau}\rho(4 k_{F},x,\tau):
\, :\partial_{\tau'}\rho(-4 k_{F},x,\tau'):
\frac{\pi v_\rc \lambda_2}{2 \omega_2} 
\re^{-\omega_2 |\tau-\tau'|}
\nonumber \\ [2mm]
& = &
-4 \pi^2 v_c a_u^2 \, \frac{\lambda_2}{\omega_2}
\p_{\tau} \Phi_c(\tau)\int \rd \tau' \re^{-\omega_2 |\tau'|} \exp\{\ri \sqrt{8
\pi} [\Phi_c(\tau)-\Phi_c(\tau-\tau')]\}
\p_\tau \Phi_c(\tau-\tau') \quad .
\nonumber
\eea
If we note with 
\be
\bar{f}_{4k_F}(\tau') = 4\pi^2 v_c a_u^2 \,  \frac{\lambda_2}{\omega_2} \, 
\re^{-\omega_2 |\tau'|} \la \re^{ \ri \sqrt{8\pi} \Phi_c(\tau)}
\re^{-\ri \sqrt{8\pi} \Phi_c(\tau-\tau')} \ra
\quad , \quad F_{4k_F}(\tau)= \int \rd \tau' \bar{f}_{4k_F}(\tau') \Phi_c(\tau-\tau') \quad ,
\ee
in the following we make the mean field approximation by
representing the $4k_F$ contribution  in $\omega$-space in the following
form
\be
\int \rd \tau \left[\p_\tau \Phi_c(\tau)\right] 
\left[\p_\tau F_{4k_F}(\tau)\right] = 
\int \rd \omega \, \omega^2 \, \Phi_c(-\omega) F_{4k_F}(\omega) =
\int \rd \omega \, \omega^2 \, \Phi_c(-\omega) \bar{f}_{4k_F}(\omega) \Phi_c(\omega)
\quad .
\label{om-space-f4F}
\ee 
The $2k_F$ components give for $\bar{f}_{2k_F}$ two terms from which, within this 
approximation, only one gives a finite contribution in the dynamics. The first one 
being 
\be
\bar{f}_{1,2k_F}(\tau') = \pi^2 v_c a_c^2 \,  \frac{\lambda_1}{\omega_1} \,
\re^{-\omega_1 |\tau'|}\,  \la\,  \re^{ \ri \sqrt{2\pi}
\Phi_c(\tau)}
\re^{-\ri \sqrt{2\pi} \Phi_c(\tau-\tau')}
\cos\left[\sqrt{2\pi}\Phi_s(\tau)\right]
\cos\left[\sqrt{2\pi}\Phi_s(\tau - \tau')\right]\,  \ra \, \quad .
\label{bar-f1-2kF}
\ee
is similar to $\bar{f}_{4k_F}$. 
The additional one is 
\be
\bar{f}_{2,2k_F}(\tau') = \pi^2 v_c  a_c^2 \, \frac{\lambda_1}{\omega_1} \,
\re^{-\omega_1 |\tau'|}\,  \la\,  \re^{ \ri \sqrt{2\pi}
\Phi_c(\tau)}
\re^{-\ri \sqrt{2\pi} \Phi_c(\tau-\tau')} \, \ra
\label{bar-f2-2kF}
\ee
multiplying
\be
\p_\tau \cos\left[\sqrt{2\pi}\Phi_s(\tau)\right]
\p_\tau\cos\left[\sqrt{2\pi}\Phi_s(\tau - \tau')\right] \quad .
\ee

In $\omega$-space the term (\ref{bar-f1-2kF}) will give a contribution similar to
(\ref{om-space-f4F}), namely
\be
\int \rd \tau \left[\p_\tau \Phi_c(\tau)\right]
\left[\p_\tau F_{1,2k_F}(\tau)\right] =
\int \rd \omega \, \omega^2 \, \Phi_c(-\omega) \bar{f}_{1,2k_F}(\omega)
\Phi_c(\omega)
\quad ,
\ee
whereas the term (\ref{bar-f2-2kF}) can be shown to be a total time dervative and therefore we
do not add it as a contribution.

In the contributing terms $\bar{f}_{4k_F}$ and $\bar{f}_{2,2k_F}$ we have made the 
mean field approximations
\be
\exp\left\{\ri \sqrt{8\pi}\left[\Phi_c(\tau)-\Phi_c(\tau-\tau')\right]\right\}\approx 
\la \exp\left\{\ri \sqrt{8\pi}\left[\Phi_c(\tau)-
\Phi_c(\tau-\tau')\right]\right\}\ra
= \left(\frac{\tau_0}{|\tau'|}\right)^{2d_{4k_F}}
\ee
and
\bea
& &\hspace{20mm}\exp\left\{\ri \sqrt{2\pi}\left[\Phi_c(\tau)-\Phi_c(\tau-\tau')\right]\right\}
\cos\left[\sqrt{2\pi}\Phi_c(\tau)\right]
\cos\left[\sqrt{2\pi}\Phi_c(\tau-\tau')\right] \approx
\nonumber \\[2mm]
& & \la\exp\left\{\ri \sqrt{2\pi}\left[\Phi_c(\tau)-
\Phi_c(\tau-\tau')\right]\right\}\ra
\la \cos\left[\sqrt{2\pi}\Phi_c(\tau)\right]
\cos\left[\sqrt{2\pi}\Phi_c(\tau-\tau')\right]\ra \sim
\left( \frac{\tau_0}{|\tau'|}\right)^{2d_{2k_F}}
\left( \frac{\tau_0}{|\tau'|}\right)^{2d_{s}} \quad ,
\eea
where $d_{4k_F}=2K_c^{0}$ and $d_{2 k_F}=K_c^{0}/2$, $d_s =1/2$ and
\be
\int_{-\infty}^{+\infty} \rd \tau' \re^{-\omega_2 |\tau'|} \left(\frac{\tau_0}
{\tau'}\right)^{4K_c^0}\approx \frac{2\tau_0}{4K_c^0-1} \quad .
\label{tau'-integr}
\ee
$\tau_0$ is a low cut-off of the order of $\epsilon_F^{-1}$. The integral
(\ref{tau'-integr}) is always positive since in the Hubbard model $K_c^{0}$ takes
values decreasing from $1$ to $1/2$ as $u$=U/t increases from $0$ to $\infty$ (see for instance 
Frahm and Korepin \cite{Frahm}).

In $\omega$-space the effect of the nonlocal term of (\ref{S_int}) is that 
the charge sector can be described by 
the modified Gaussian action with the kinetic energy nonlocal in time:
\begin{equation}
S=\frac{1}{2K_c^{0}} \sum_{\omega ,q}\Phi_{\rm c}(-\omega,-q)
\left[\frac{1}{v_{\rm c}}
\omega^2 f(\omega)+v_{\rm c} q^2\right] \Phi_{\rm c}(\omega,q)
\quad , \quad f(\omega)=
1+v_{\rm c} K_c^{0} \Bigl[\bar{f}_{1,2k_F}(\omega) + \bar{f}_{4k_F}(\omega) \Bigr] \quad ,
\label{disp}
\end{equation}
where the function $f(\omega)$  
takes values between 
\be
f(\omega \gg 0)\, := 1 + v_{\rm c} K_c^{0}\int \rd \tau' \left[\bar{f}_{1,2k_F}(\tau') + 
\bar{f}_{4k_F}(\tau') \right] = 1 + 
2\pi v_c^2 K_c^{0} \tau_0 
\left\{a_c^2 \frac{\lambda_1}{\omega_1}
+4a_u^2 \frac{\lambda_2K_c^{0}}{\omega_2(4 K_c^{0}-1)}\right\}= k^2 
\label{k^2}
\ee 
and $f(\omega \gg 0) \, :=1$. In Eq.~(\ref{k^2}) we introduced the $\omega$-independent 
constant $k$.

It is interesting to explore the consequences of frequency dependence of $f(\omega)$. 
Its form, within our approximation, is given by
\begin{equation}
f(\omega)=1+\frac{\omega_{\rm 0}^2}{\omega^2+\omega_{\rm 0}^2}(k^2-1)
\quad , 
\end{equation}
where $f(0)=k^2$ (the parameter $k$ takes large values) and 
$f(\omega \gg 0)=1$. $\omega_0$ is considered to be the phonon's frequency (the phonons are 
considered as without dispersion and we can approximate 
$\omega_{2k_F}\approx \omega_{4k_F}=\omega_0$).
Large values of $k$ notify strong renormalization, and this is stronger at
smaller frequencies (but not smaller than the critical one set from the spin sectors 
requirement of no gap opening). The value of the constant $k$ is expected to be big, and 
experimental observations in materials like ${\rm K_{0.3}MoO_3}$ \cite{Grunner} give a value of 
several hundred.

In time representation the action on the charge sector has the following form
\bea
S & = & \int {\rm d} x {\rm d} \tau \left\{
\frac{1}{2K_c^{0}}\left[
\frac{1}{v_c}
\left(\partial_\tau \Phi_{\rc}\right)^2
+v_\rc \left(\partial_x \Phi_{\rc}\right)^2\right]
+ 
\partial_\tau \Phi_c(\tau) \int \rd \tau'
\left[\bar{f}_{1,2k_F}(\tau') + \bar{f}_{4k_F}(\tau') \right]
\p_\tau \Phi_c(\tau-\tau') \right \} \quad .
\label{renorm-action}
\eea
The effective retarded attraction between the staggered components of the
charge density affects only the charge sector.  In the spin sector $K_s^{0}$
remains $1$, [preserving the SU(2) symmetry].

For the special limit $\omega = 0$ in $f(\omega)$ of (\ref{disp}), it takes the 
following form 
\be
S = \frac{1}{2 K_c^{0}}\int {\rm d} x {\rm d} \tau \left[ \frac{1}{v_c}
k^2
\left(\partial_\tau \Phi_{\rc}\right)^2
+v_\rc \left(\partial_x \Phi_{\rc}\right)^2\right] \quad , \quad 
[f(\omega = 0)] \quad , 
\label{renorm-action1}
\ee
which in the canonical form (\ref{canonical}) is described by the new renormalized
parameters
\be
K_c = 
\frac{1}{k}K_c^{0} \quad , \quad
\tilde{v}_c = 
\frac{1}{k}  v_c \quad .
\ee
The limit $\omega \gg 0$ in the charge sector is described by the action
\be
S = \frac{1}{2 K_c^{0}}\int {\rm d} x {\rm d} \tau \left[ \frac{1}{v_c}
\left(\partial_\tau \Phi_{\rc}\right)^2
+v_\rc \left(\partial_x \Phi_{\rc}\right)^2\right] \quad ,  \quad
[f(\omega \gg 0)] \quad .
\label{renorm-action2}
\ee
We consider the case of $K_c^{0}$ being near $1$ and neglect the
renormalization coming form the local term, which is of the same magnitude as the 
renormalization of the spin sector and which we considered to be small. In this
case $K_c=1/k$.

 The single-particle Green's function is expressed
as a correlator of the chiral exponents of the charge and spin fields:
\bea
G(\tau,x) &=& \la \exp\left[\ri\sqrt{2\pi}\phi_c(\tau,x)\right]
\exp\left[\ri\sqrt{2\pi} \phi_s(\tau,x)\right]\cdot
\exp\left[\ri\sqrt{2\pi}\phi_c(0,0)\right]
\exp\left[\ri\sqrt{2\pi}
\phi_s(0,0)\right] \ra
\nonumber\\ [2mm]
& = & (\tau - x/v_s)^{-1/2}\la \exp\left[\ri\sqrt{2\pi}\phi_c(\tau,x)\right]
\exp\left[\ri\sqrt{2\pi}\phi_c(0,0)\right]\ra
\eea

The model (\ref{disp}) leads to the Matsubara Green's function
\bea
G(x=0;\tau)&=&\frac{1}{\tau}\exp\left\{-\frac{1}{2}\int_{0}^{
\epsilon_F/\omega_0}
\frac{{\rm d}x}{x}\left[\left(\frac{x^2+1}{x^2+k^2}\right)^{1/4}-
\left(\frac{x^2+k^2}{x^2+1}\right)^{1/4} \right]^2\sin^2\left(\omega_0 \tau
x/2\right) \right\}
\quad .
\label{time-mats}
\eea

This Green's function for large $\tau$ behaves as
\be
\label{green-funct}
G(\tau) \stackrel{(\tau \rightarrow +\infty)}{\sim} \frac{1}{\tau^{\theta_c+1}} \quad 
\Longrightarrow \quad  
\rho_{_{{\small {\rm DOS}}}}(\omega) \stackrel{(\omega \rightarrow 0^+)}{\sim}
 \omega^\theta_c
\quad  ; \quad \theta_c=
\frac{1}{4}\left(\frac{1}{\sqrt{K_c}} - \sqrt{K_c}\right)^2 \quad ; \quad 
[f(\omega=0)] \quad , 
\ee
which corresponds to $f(\omega=0)$, Eq. (\ref{renorm-action1}).
This is so because the exponential of the Green's function Eq.~(\ref{green-funct}) 
for large $\tau$ behaves as $-[(\sqrt{k}-1/\sqrt{k})^2/4]\ln \tau$.

For small $\tau$, it bahaves as
\be
G(\tau) \sim \frac{1}{\tau} \quad \Longrightarrow \quad
\rho_{_{{\small {\rm DOS}}}}(\omega) \stackrel{(\omega \gg 0)}{\longrightarrow} 
{\rm const}  \quad , \quad 
(\tau \rightarrow 0) \quad ; 
\quad  [f(\omega \gg 0)] \quad ,
\ee
corresponding to the case with $f(\omega)$ with large $\omega$, 
Eq.~(\ref{renorm-action2}), [if $G(\tau)\sim 1/\tau$ were in all $\tau$ space then
$\tilde{G}(i \omega_n)=-2i\int_0^{+\infty}\rd \tau \sin(\omega_n \tau)/\tau= -2 i \,
{\rm Si}(+\infty)=-i \pi$ and $\rho_{_{{\small {\rm DOS}}}}(\omega)=(-1/\pi)
\Im {\rm m} G^R=1$].

To extract DOS on the whole frequency space one has to Fourier transform this
function to get $\tilde{G}(i\omega_{\rm n})$ and then take the analytic
continuation $i \omega_n \rightarrow \omega +i 0$.
The values of $\rho_{_{{\small {\rm DOS}}}}(\omega)$ between the two asymptotic
limits, $\omega \rightarrow 0^+$ and $\omega \gg 0$, are shown
in Fig~1.
They are obtained numerically
by analytic continuation of $\tilde{G}(i\omega_{\rm n})$ from the imaginary axis
to just above the real axis.
A remarkable feature of 
DOS is the peak at $\omega \approx K^{-1}\omega_0/2$ (see Fig. 2). 

Since DOS
is invariant under $K \rightarrow K^{-1}$ this empirical formula can
be generalized as
\begin{equation}
\omega_{\rm peak} = \frac{1}{2}(K^{-1} + K)\omega_0
\label{peak}
\quad .
\end{equation}
It  is interesting
that the peak always
occurs at frequencies larger than the characteristic
phonon frequency
$\omega_0$. At small (large) values of $K$ this discrepancy can be
quite substantial. For example, a  peak in DOS
has been observed in K$_{0.3}$MoO$_3$ at
$\omega \approx 300$ meV. The behavior of DOS at smaller frequencies
is almost linear in $\omega$ which suggests $K \approx 0.15$. Then
Eq. (\ref{peak}) gives  a
reasonable estimate for the phonon frequency: $\omega_0 \approx  90$ meV.

For smaller $\omega_0$ the Luttinger parameter $K_c$ becomes small, the
exponent $\theta_c$ becomes large and the peak approaches the origin. This 
is also observed in the numerical plots, and is in accord with (\ref{peak}).

The Fig.~1 clearly show also the crossover from the Luttinger-liquid type
behavior with singular $d\rho/d\omega$ at $K >  0.17$ to the
power-low suppressed with exponent larger than 1 behavior at $K < 0.17$.

To summarize, we have investigated in this paper the single particle density of states at the 
fermi level of 1D conductors at incommensurate fillings with electron-phonon interactions. 
The problem is studied using the Tomonaga-Luttinger liquid approch. The phonon degrees of freedom 
were integrated out and the remaining effective electron-electron interactions were studied. 
The electron-phonon interactions with large momentum transfer of $q\approx 2k_F$ and 
$q\approx 4k_F$ result in retardation effects which lead to the unusual form of LL model, namely 
with parameters of the charge sector depending on the frequency. 
The LL constant $K$ is also a function of $\omega_0$, the phonon frequency.
At small frequencies 
$K_c$ and $v_c$ are strongly renormalized by the interactions.
We found the imaginary time Matsubara Green's function for this model and built the single particle density of states. 
Dos exhibits a peak which is proportional with the phonon frequency,  but nevertheless
they do not coincide.

E.P. is grateful to Steve Allen and to Dave Allen for helpful
discussions  and to A.~M. Tsvelik for suggesting the problem and discussions.

\end{document}